# Tensor Decomposition based Adaptive Model Reduction for Power System Simulation

Denis Osipov, *Student Member, IEEE*, and Kai Sun, *Senior Member, IEEE*

*Abstract*—The letter proposes an adaptive model reduction approach based on tensor decomposition to speed up time-domain power system simulation. Taylor series expansion of a power system dynamic model is calculated around multiple equilibria corresponding to different load levels. The terms of Taylor expansion are converted to the tensor format and reduced into smaller-size matrices with the help of tensor decomposition. The approach adaptively changes the complexity of a power system model based on the size of a disturbance to maintain the compromise between high simulation speed and high accuracy of the reduced model. The proposed approach is compared with a traditional linear model reduction approach on the 140-bus 48-machine Northeast Power Coordinating Council system.

*Index Terms*--Model reduction, power system, simulation speed, Taylor series expansion, tensor decomposition.

## I. INTRODUCTION

FAST power system simulation is valuable for online transient stability assessment to predict potential instability following a disturbance. Knowledge of potential system instabilities in real time is important as it allows a system operator to perform timely control actions to save the system. One way to increase the speed of online simulation is to apply model reduction to a power system.

A widely adopted model reduction approach for a large-scale power system is to divide the system into two areas: 1) the study area, i.e. the focus of simulation and stability assessment, where all details of models are preserved, and from where all disturbances are originated; 2) the external area, where models can be approximated to improve speed of the whole system simulation. Traditionally the external system is reduced using coherency-based methods [1] or linear model reduction methods such as balanced truncation [2], Krylov subspace [3], dominant pole [4], low-rank Choleski factors [5] based methods. These techniques perform well when the size of a disturbance is small. However, when a large disturbance happens the model reduction error can increase to an unsatisfactory large value. Nonlinear model reduction can improve the accuracy; however, as shown in [6] the speed performance gain from the reduction is substantially decreased. This work proposes to use tensor decomposition to represent the Taylor series expansion of a power system dynamic model in order to further improve the accuracy of the adaptive model reduction proposed in [7] while maintaining high simulation speed.

## II. TENSOR DECOMPOSITION BASED MODEL REDUCTION

### A. Power System Model Approximation

In this work, each generator of the power system is represented by a two-axis model with a non-reheat steam turbine model, a first-order governor model and an IEEE type-1 exciter model as described by (1), which contains nine differential equations given in [7].

$$\begin{cases} \dot{\mathbf{x}} = \mathbf{f}(\mathbf{x}) \\ \mathbf{y} = \mathbf{x} \end{cases} \quad (1)$$

where $\mathbf{x} \in R^n$ is the state vector, $\mathbf{y} \in R^n$ is the output vector depending on the required simulation results, which is typically equal to the state vector or a portion of it, and $n$ is the total number of state variables.

The system in (1) can be approximated by Taylor series expansion as shown below in the matrix formulation:

$$\begin{cases} \Delta\dot{\mathbf{x}} = \mathbf{A}_1\Delta\mathbf{x} + \mathbf{A}_2(\Delta\mathbf{x} \otimes \Delta\mathbf{x}) + \mathbf{A}_3(\Delta\mathbf{x} \otimes \Delta\mathbf{x} \otimes \Delta\mathbf{x}) + \cdots \\ \mathbf{y} = \Delta\mathbf{x} + \mathbf{x}_0 \end{cases} \quad (2)$$

where $\mathbf{x}_0$ is the initial state vector; $\mathbf{A}_i \in R^{n \times n^i}$ is the matrix of partial derivatives of the functions in (1) of order $i$; $\Delta\mathbf{x}$ is the deviation vector of state variables; "$\otimes$" denotes Kronecker product.

In system (2), the dimensions of matrices $\mathbf{A}_i$ grow exponentially with the order, which in turn increases the computational burden and can make the approximated model in (2) even slower than the original model in (1). To address this, we propose to represent matrices $\mathbf{A}_i$ as tensors and apply tensor decomposition to decrease the size of the matrices and improve the speed of computation.

### B. Tensor representation

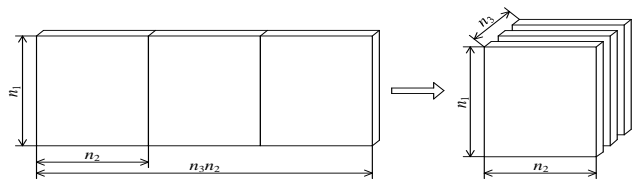

Fig. 1. Representation of a matrix by a third-order tensor.

A tensor is a multidimensional array that is defined as

---

This work was supported in part by the ERC Program of the NSF and DOE under grant EEC-1041877 and in part by the NSF grant ECCS-1610025.

D. Osipov and K. Sun are with the Department of Electrical Engineering and Computer Science, University of Tennessee, Knoxville, TN 37996 USA (email: dosipov@vols.utk.edu, kaisun@utk.edu).

$\mathcal{A} \in R^{n_1 \times n_2 \times \cdots \times n_k \times \cdots \times n_d}$, where $n_k$ is the size in dimension $k$, $d$ is the number of dimensions. A matrix can be converted to a tensor as shown in the example in Fig. 1.

Matrices $\mathbf{A}_2, \mathbf{A}_3, \ldots$ in (2) can be converted to tensors $\mathcal{A}_2, \mathcal{A}_3, \ldots$, where $\mathcal{A}_2 \in R^{n_1 \times n_2 \times n_3}$, $\mathcal{A}_3 \in R^{n_1 \times n_2 \times n_3 \times n_4}$, $n_1=n_2=n_3=n_4=n$. Kronecker product in (2) can be represented by the tensor dimension multiplication [8]. A $k$-dimension product of a tensor and a matrix is a tensor of which the entries are calculated as follows [9]:

$$(\mathcal{A} \times_k X)_{j_1 j_2 \cdots j_{k-1} i j_{k+1} \cdots j_d} \stackrel{\text{def}}{=} \sum_{j_k=1}^{n_k} \mathcal{A}_{j_1 j_2 \cdots j_k \cdots j_d} X_{i j_k} \quad (3)$$

where $\mathcal{A} \times_k X \in R^{n_1 \times n_2 \times \cdots \times n_{k-1} \times m \times n_{k+1} \times \cdots \times n_d}$, $X \in R^{m \times n_k}$.

The system (2) can be rewritten as:

$$\begin{cases} \Delta \dot{\mathbf{x}} = \mathbf{A}_1 \Delta \mathbf{x} + \mathcal{A}_2 \times_2 \Delta \mathbf{x}^T \times_3 \Delta \mathbf{x}^T + \mathcal{A}_3 \times_2 \Delta \mathbf{x}^T \times_3 \Delta \mathbf{x}^T \times_4 \Delta \mathbf{x}^T + \cdots \\ \mathbf{y} = \Delta \mathbf{x} + \mathbf{x}_0 \end{cases} \quad (4)$$

*C. Tensor decomposition*

Using the CANDECOMP/PARAFAC (CP) decomposition [10], a tensor can be approximated by the sum of a finite number of rank-one tensors. A $d$th-order tensor is rank one if it can be written as the outer product of $d$ vectors: $\mathcal{A} = a^{(1)} \circ a^{(2)} \circ \cdots \circ a^{(k)} \circ \cdots \circ a^{(d)}$, where $a^{(k)} \in R^{n_k}$ is the $k$th rank-one component. CP tensor decomposition can be written as

$$\mathcal{A} \approx \sum_{i=1}^{r} a_i^{(1)} \circ a_i^{(2)} \circ \cdots \circ a_i^{(d)}, \quad (5)$$

where $r$ is the rank of the decomposed tensor. An example of CP tensor decomposition is illustrated in Fig. 2.

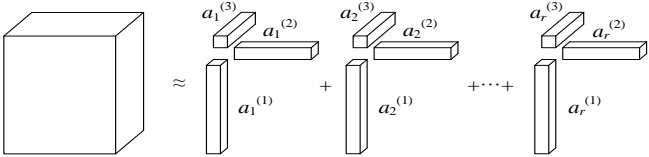

Fig. 2. CP decomposition of a third-order tensor.

Components in (5) which correspond to the same dimension can be grouped into a factor matrix $\mathbf{A}^{(k)} = [a_1^{(k)}, a_2^{(k)}, \ldots, a_r^{(k)}]$, where $\mathbf{A}^{(k)} \in R^{n_k \times r}$.

Factor matrices can be used in one-dimension matricization of a tensor to reconstruct the original matrix [10]:

$$\mathbf{A} \approx \mathbf{A}^{(1)} \left( \mathbf{A}^{(d)} \odot \cdots \odot \mathbf{A}^{(3)} \odot \mathbf{A}^{(2)} \right)^T, \quad (6)$$

where $\odot$ denotes Khatri-Rao product.

Tensors in (4) are matricized with the help of (6) and the following system is obtained:

$$\begin{cases} \Delta \dot{\mathbf{x}} = \mathbf{A}_1 \Delta \mathbf{x} + \mathbf{A}_2^{(1)} \left( \Delta \mathbf{x}^T \mathbf{A}_2^{(3)} \odot \Delta \mathbf{x}^T \mathbf{A}_2^{(2)} \right)^T \\ \quad + \mathbf{A}_3^{(1)} \left( \Delta \mathbf{x}^T \mathbf{A}_3^{(4)} \odot \Delta \mathbf{x}^T \mathbf{A}_3^{(3)} \odot \Delta \mathbf{x}^T \mathbf{A}_3^{(2)} \right)^T + \cdots \\ \mathbf{y} = \Delta \mathbf{x} + \mathbf{x}_0 \end{cases} \quad (7)$$

*D. Proposed Hybrid Model Reduction*

The proposed model reduction approach first partitions the system into two areas: the study area and the external area. All generators of the study area and the generators of the external area electrically close to the boundary between the areas are described by the original nonlinear equations. Then, the model of each external generator is reduced by approximation using the Taylor expansion series up to a certain order, which is represented through the tensor decomposition. The resulting hybrid system is described by the following expression:

$$\begin{cases} \Delta \dot{\mathbf{x}} = \begin{pmatrix} \mathbf{A}_1 \Delta \mathbf{x} + \mathbf{A}_2^{(1)} \left( \Delta \mathbf{x}^T \mathbf{A}_2^{(3)} \odot \Delta \mathbf{x}^T \mathbf{A}_2^{(2)} \right)^T + \mathbf{A}_3^{(1)} \left( \Delta \mathbf{x}^T \mathbf{A}_3^{(4)} \odot \Delta \mathbf{x}^T \mathbf{A}_3^{(3)} \odot \Delta \mathbf{x}^T \mathbf{A}_3^{(2)} \right)^T + \cdots \\ \hat{\mathbf{f}}(\mathbf{x}) - \mathbf{x}_0 \end{pmatrix} \\ \mathbf{y} = \Delta \mathbf{x} + \mathbf{x}_0 \end{cases} \quad (8)$$

where $\hat{\mathbf{f}}$ comprises the functions that are kept nonlinear. The selection of generators is performed based on the column norm of the admittance matrix as introduced in [7].

*E. Adaptive Switching Algorithm*

As the requirements for the details of the simulated system depend on the severity of a contingency, this work proposes this adaptive algorithm: 1) use the original system (1) during the fault-on condition; 2) in the post-fault condition, use the hybrid system (8) if the disturbance is large, or otherwise, use the Taylor series expansion based system (7).

The size of the disturbance is determined by the maximum rotor angle deviation of all generators of the study area. A generator with large inertia located electrically far away from the boundary is selected as the reference generator.

During the simulation the algorithm checks if there is a large change in the system load level. If the load level changes by more than 10% the tensor decomposition matrices are chosen from the set of matrices calculated in advance off-line to correspond to the new load level. This allows the algorithm to maintain the accuracy after a large operating condition change and differentiate the algorithm from the one described in [7]. Another difference compared to the switching algorithm in [7] is the use of Taylor series expansion in the form of tensor decomposition instead of a linearized system and as a part of the hybrid system (8) which is a combination of Taylor series expansion in the form of tensor decomposition and nonlinear functions instead of a combination of linear and nonlinear functions in [7].

## III. CASE STUDIES

The proposed approach is tested on a 140-bus 48-machine Northeast Power Coordinating Council system [11]. The study area is defined as New England part of the system with 36 buses, 9 generators and 9×9=81 state variables. The external area is defined as the rest of the system with 104 buses, 39 generators and 39×9=351 state variables.

The threshold for the column norm of the admittance matrix that determines whether a generator is electrically close to the boundary between two areas is set to 1 p.u. based on the case study in [7]. This corresponds to the approximation of 34 out of 48 generators by Taylor series. The expansion is performed up to the 3$^{\text{rd}}$ order and converted to the tensor form. Tensor decomposition is computed with ranks 27 and 29 respectively for the 2$^{\text{nd}}$ and 3$^{\text{rd}}$ order terms.

The ranks are selected in a case study where the rank is increased from 1 until the increase in rank does not improve the accuracy of the approach in terms of rotor angle by at least 0.1 degree. Another case study is conducted to set the threshold for the maximum rotor angle deviation that controls the switching between the model with tensor decomposition and the hybrid model. The threshold is increased from 1 degree until the largest rotor angle error for all generators in the study area is below 5 degrees, which turns out a the threshold at 26 degrees. The simulations are performed in MATLAB R2015a on a computer with the 4-GHz AMD FX-8350 processor. The simulation length and integration step are respectively set to 16 seconds and 0.01 second.

In a case study of all different faults, the values of critical clearing time (CCT) in the system reduced with proposed approach are identical to the values obtained from simulation of the original system. Thus, the proposed approach maintains stability of the reduced system when the original system is stable. The generator with the largest rotor angle error following a fault at the bus with the longest CCT is used to compare the proposed approach with the traditional linear model reduction approach described in [7]. The results of the comparison are shown in Fig. 3. The linear model reduction generates a large error while the trajectory simulated from the proposed approach closely follows that of the original model.

For a quantitative comparison, the root mean square (RMS) error of the rotor angle is calculated, which equals 22.4 degrees with the linear model reduction and equals 4.3 degrees (reduced by 81%) with the proposed approach.

In addition, the proposed approach is tested in terms of speed performance. Table I compares the simulation times respectively with the original model, with linear model reduction, the adaptive model reduction in [7] and the proposed adaptive model reduction. As the tensor decomposition is performed off-line, the calculation time for the reduced model matrices is not included in Table I. The proposed approach reduces the simulation time by 76% compare to that using the original system. The speed performance of the adaptive model reduction is identical to the traditional linear model reduction approach while the accuracy of the simulation is substantially higher. The proposed approach enables faster simulation than the adaptive approach in [7] because the use of higher-order terms of Taylor series allows a larger threshold for the rotor angle deviation and thus earlier switching to a faster model (i.e. from the hybrid model to the Taylor series only model).

The RMS error of the model reduction described in [7] is 5.1 degrees. Thus, the proposed tensor decomposition based adaptive model reduction reduces the simulation time and improves the accuracy of the simulation.

To test how the proposed approach performs with variations of the operating condition, the original load level is increased and decreased at 5% increments to create totally 9 conditions. The fault is set at bus 3 with the duration equal to CCT. The load levels of 80%, 100% and 120% are selected as 3 representative levels to perform the proposed approach and each covers 3 of 9 conditions as shown in Table II. When the load level changes by more than 10%, tensor decomposition matrices are changed to the ones obtained from the Taylor expansion calculated around the representative condition with a higher or lower load level depending on the direction of load change. Table II gives the CCT of the fault for each condition and the rotor angle RMS error in simulating each of 9 conditions using the reduced model on its representative condition. All RMS errors are within 5 degrees. Thus, the adaptive model reduction is capable of accurate system representation at moderately different operating conditions.

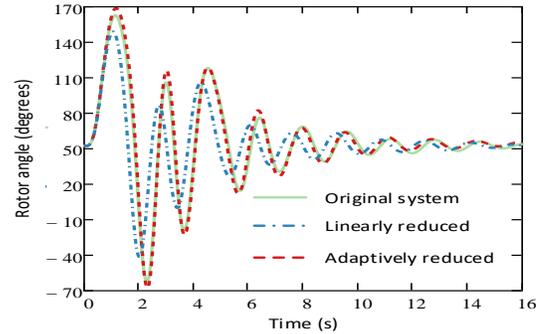

Fig. 3. Rotor angle of generator 23 following the fault at bus 3.

TABLE I
COMPARISON OF SIMULATION TIME

| Systems | Time costs (s) |
|---|---|
| Original model | 3.7 |
| With linear model reduction | 0.9 |
| With adaptive model reduction | 1.0 |
| With tensor decomposition based adaptive model reduction | 0.9 |

TABLE II
COMPARISON OF ROTOR ANGLE RMS ERRORS AT DIFFERENT LOAD LEVELS

| Load level (%) | 120 | 115 | 110 | 105 | 100 | 95 | 90 | 85 | 80 |
|---|---|---|---|---|---|---|---|---|---|
| CCT (s) | 0.19 | 0.23 | 0.27 | 0.33 | 0.39 | 0.44 | 0.51 | 0.59 | 0.7 |
| Error (°) | 4.1 | 3.6 | 4.9 | 4.9 | 4.3 | 3.5 | 3.5 | 3.6 | 4.9 |

## IV. CONCLUSIONS

The proposed tensor decomposition based adaptive model reduction approach improves the speed of power system transient stability simulation while maintaining a satisfactory level of accuracy.

## V. REFERENCES

[1] J. H. Chow, *Power system coherency and model reduction*. Springer, 2013, pp. 1-3.
[2] S. Liu, "Dynamic-data driven real-time identification for electric power systems," Ph.D. dissertation, UIUC, Urbana, IL, 2009.
[3] D. Chaniotis, "Krylov subspace methods in power system studies," Ph.D. dissertation, UIUC, Urbana, IL, 2001.
[4] J. Rommes, "Modal Approximation and Computation of Dominant Poles," in *Model Order Reduction: Theory, Research Aspects and Applications*, Springer, 2008, pp. 177–193.
[5] F. D. Freitas, J. Rommes, N. Martins, "Gramian-Based Reduction Method Applied to Large Sparse Power System Descriptor Models," *IEEE Trans. Power Syst., vol. 23, no. 3, pp. 1258–1270, Aug. 2008.*
[6] M. Striebel, J. Rommes, "Model order reduction of nonlinear systems in circuit simulation: status and applications," in *Model Reduction for Circuit Simulation*, Springer, 2011, pp. 289–301.
[7] D. Osipov, K. Sun, "Adaptive nonlinear model reduction for fast power system simulation," *IEEE Trans. Power Syst., vol. 33, no. 6, pp. 6746–6754, Nov. 2018.*
[8] T. G. Kolda, (2006, Apr.). Multilinear operators for higher-order decompositions. Sandia National Laboratories, Livermore, CA.
[9] L. de Lathauwer, et al, "A multilinear singular value decomposition," *SIAM Review*, vol. 21, no. 4, pp. 1253–1278, Aug. 2000.
[10] T. Kolda, B. Bader, "Tensor decomposition and applications," *SIAM Review*, vol. 51, no. 3, pp. 455–500, Aug. 2009.
[11] J.H. Chow, K.W. Cheung, "A toolbox for power system dynamics and control engineering education and research," IEEE Trans. Power Syst., vol.7, no.4, pp.1559-1564, Nov 1992.